\documentclass[12pt]{spieman}  % 12pt font required by SPIE;
\usepackage{amsmath,amsfonts,amssymb}
\usepackage{graphicx}
\usepackage{setspace}
\usepackage{tocloft}
\usepackage[flushleft]{threeparttable}
\usepackage{lineno}
\usepackage{aas_macros}
\DeclareUnicodeCharacter{2243}{\simeq}

\title{A Census of the Most Obscured Galaxy Nuclei over Cosmic Time to be revealed by PRIMA}

\author[a, *]{Fergus R. Donnan}
\author[a]{Dimitra Rigopoulou}
\author[b]{Ismael Garc{\'i}a-Bernete}
\author[c]{Laura Bisigello}
\author[d]{Susanne Aalto}

\affil[a]{Department of Physics, University of Oxford, Keble Road, Oxford, OX1 3RH, UK}
\affil[b]{Centro de Astrobiolog\'{\i}a (CAB), CSIC-INTA, Camino Bajo del Castillo s/n, E-28692 Villanueva de la Ca\~nada, Madrid, Spain}
\affil[c]{INAF, Osservatorio Astronomico di Padova, Vicolo, dell'Osservatorio 5, I-35122, Padova, Italy}
\affil[d]{Department of Space, Earth and Environment, Chalmers University of Technology, SE-412 96, Gothenburg, Sweden}

\cftpagenumbersoff{figure}
\cftpagenumbersoff{table} 
\begin{document} 
\maketitle

\begin{abstract}
Characterizing the growth of supermassive Black Holes (SMBHs) is critical to the evolution of galaxies, however the majority of this activity is obscured, rendering traditional tracers of active SMBHs, such as in the restframe optical/UV, ineffective. The mid-infrared has been particularly successful in revealing obscured AGN activity however much of this work is confined to the local universe due to the lack of a far-IR telescope with the required sensitivity and wavelength coverage.
In this work we demonstrate the effectiveness of PRIMA (PRobe far-Infrared Mission for Astrophysics), a concept 1.8m far-IR observatory, to detect and characterize deeply obscured galaxy nuclei over cosmic time. With the PRIMAger instrument covering 25 - 235 $\mu$m, we find that we can accurately detect obscured nuclei via the deep silicate absorption at restframe $9.8 \mu$m between $z=2-7$. Additionally, the FIRESS spectrograph can produce R$\sim$100 spectra of obscured nuclei out to $z\sim7$, detecting Polycyclic Aromatic Hydrocarbons (PAHs), ices, ionized and molecular gas. With the large number of deeply obscured nuclei PRIMA can detect and characterize, such a mission is critical to understanding the growth of SMBHs. 

\end{abstract}

% Include a list of up to six keywords after the abstract
\keywords{PRIMA, far-infrared, space telescopes, galaxy evolution}

% Include email contact information for corresponding author
{\noindent \footnotesize\textbf{*}Fergus R. Donnan,  \linkable{fergus.donnan@physics.ox.ac.uk} }

\begin{spacing}{2}   % use double spacing for rest of manuscrip

% Intro

% Generate example spectra 

% Estimate how many objects we can detect as a function of redshift

\section{Introduction}
\label{sect:intro}  % \label{} allows reference to this section

% Introduce role of AGN/obscuration, x-ray background, CONs etc

Active Galactic Nuclei (AGN) are thought to play a key role in how galaxies evolve, where the interaction with the host galaxy can regulate star-formation  on a variety of scales\cite{Heckman2014, Combes2017}. To understand this co-evolution, one must understand how supermassive Black Holes (SMBHs) grow and affect their host galaxy, however much of this activity is hidden behind vast quantities of dust \cite{Hickox2018}. In the unified model, obscuration was thought to be a viewing angle effect where type 2 AGN are viewed through a dust obscuring torus \cite{Antonucci1993}. However this picture has evolved in recent times where obscuration may be an evolutionary stage, where rapid SMBH growth is hidden before the AGN eventually clears its dust \cite{Hopkins2006, Fabian2012}. Evidence for a population of dust-enshrouded AGN can be found in the cosmic X-ray background \cite{Ananna2019}, which postulates the presence of a large fraction of sources with column densities in excess of $10^{22}$cm$^{-2}$. 
At the extreme end of this population with $N_{\mathrm H}>10^{25}$ cm$^{-2}$, a significant fraction of high-energy photons can be absorbed, making even X-rays ineffective \cite{Ricci2021}.

% SMBH growth at high-z, to date based in broad lines which make a very small sub sample of the total population
As the obscuring dust is heated and reprocesses photons from the accretion disk, the restframe infrared is extremely bright in deeply obscured objects. With the advent of the JWST/MIRI \cite{Wright2023}, imaging has allowed the selection of AGN at high-$z$ \cite{Yang2023}, via the hot dust continuum which appears in the near-infrared, finding significantly more objects than X-ray selected AGN.  However the wavelength coverage of JWST is not sufficient to observe the restframe mid-infrared at redshifts beyond cosmic noon ($z>3$). Additionally, we can only detect objects with an extremely hot dust component as this is required to produce a strong dust continuum in the near-infrared and so most obscured targets, which typically do not have such hot dust, are therefore missed. To unveil deeply obscured nuclei sources at high-$z$ ($z>3$), access to the rest-frame $\sim5-30\mu$m regime is required.

% Difficulty detecting CONs -> need for far-IR
In the local universe, the most obscured nuclei exhibit column densities of $N_{\mathrm H}>10^{25}$ cm$^{-2}$. These so called Compact Obscured Nuclei (CONs) have been identified via dense gas tracers in the sub-mm, namely HCN-vib emission \cite{Sakamoto2010, Aalto2019, Falstad2021}. Unlike typical type 2 AGN, CONs show a significantly deeper silicate feature and a colder dust continuum \cite{Donnan2022, Garcia-Bernete2022} in the mid-infrared. Moreover, these objects are more common than one may think considering their extreme nature, comprising $\sim40\%$ of local ULIRGs \cite{Falstad2021, Donnan2022} and considering the increase in luminous infrared galaxies at higher redshifts \cite{Zavala2021}, these objects are likely numerous.
While detecting HCN-vib in a sample of local targets has successfully identified CONs \cite{Falstad2021}, this tracer is extremely faint, making finding CONs at high-$z$ a challenge.

% Introduce mid-IR selection method 
The mid-infrared continuum of deeply obscured nuclei exhibits deep silicate absorption at $\sim 9.8\mu$m, which can be used to identify such targets \cite{Spoon2007}. A new method to select deeply obscured nuclei based on the ratio of the equivalent widths (EWs) of PAH features within the silicate band \cite{Garcia-Bernete2022} has proven to be very successful as it is less affected by contamination by the host galaxy. The method relies on the contrast between the extinction of the PAHs (tracing the circumnuclear star-formation) and the continuum (resulting from the obscured nucleus), successfully recovering the CONs identified in the sub-mm \cite{Garcia-Bernete2022, Donnan2022}. Additionally, ice absorption features are present in the spectra \cite{IDEOS} and can be used to measure the obscuring column density \cite{Garcia-Bernete2024a} based on their optical depth.

Higher ionization potential lines such as [Ne V] and [O IV] have been proposed to measure the black hole accretion rate  \cite{Stone2022}, however this method becomes ineffective for deeply obscured nuclei where these lines are typically undetected. In such deeply obscured objects, the dust absorbs the ionizing photons, preventing the gas from being ionized \cite{Abel2009, Fischer2014}. This could explain the absence of high ionization lines in the spectra of deeply obscured nuclei \cite{Donnan2022b, Perna2024}, even with the sensitivity of JWST/MIRI in the local universe. The restframe mid-infrared continuum is therefore key to unveiling the most obscured nuclei across cosmic time, which requires a far-infrared telescope that is sensitive enough to detect these objects.

% Torus modelling

%Introdcue PRIMA
PRIMA \footnote{\url{https://prima.ipac.caltech.edu/}} (PRobe far-Infrared Mission for Astrophysics) \cite{Moullet2023, Rodgers2023} is a concept for a 1.8 m cryogenically cooled FIR telescope which has recently been selected for Phase A study by NASA. The telescope will contain two instruments, PRIMAger and FIRESS. The latter is a spectrometer between 24 and 235 $\mu$m while the former is an imager. With FIRESS, there are two modes a high sensitivity low-resolution mode at R $\sim$100 while the high-resolution FTS mode allows R $\sim$ 20000 at 25 $\mu$m to R $\sim$ 4400 at 112 $\mu$m.

In this paper we demonstrate PRIMA's ability to identify deeply obscured nuclei, which have a deep silicate feature indicative of column densities of $N_{\mathrm H}\gtrsim10^{25}$ cm$^{-2}$ that have potentially remained undetected through recent/current X-ray or near-infrared missions.  We show that the identification of this population is now possible through imaging and spectroscopy with PRIMA. 

In Section \ref{sec:PRIMAgerResults} we demonstrate PRIMAger's capability of finding candidates with both real data in the local universe and simulated observations. In Section \ref{sec:FIRESS} we discuss the use of FIRESS to study the nature and evolution of these sources. In this work we assume $\Lambda$CDM cosmology with $H_0 = 70$ km s$^{-1}$ Mpc$^{-1}$, $\Omega_m = 0.27$, $\Omega_{\Lambda} = 0.73$.

% \begin{table}[ht]
% \caption{Fonts sizes and styles.} 
% \label{tab:fonts}
% \begin{center}       
% \begin{tabular}{|l|l|} %% this creates two columns
% %% |l|l| to left justify each column entry
% %% |c|c| to center each column entry
% %% use of \rule[]{}{} below opens up each row
% \hline
% \rule[-1ex]{0pt}{3.5ex}  Document entity & Brief description  \\
% \hline\hline
% \rule[-1ex]{0pt}{3.5ex}  Article title & 16 pt., bold, left justified  \\
% \hline
% \rule[-1ex]{0pt}{3.5ex}  Author names & 12 pt., bold, left justified   \\
% \hline
% \rule[-1ex]{0pt}{3.5ex}  Author affiliations & 10 pt., left justified   \\
% \hline
% \rule[-1ex]{0pt}{3.5ex}  Abstract & 10 pt.  \\
% \hline
% \rule[-1ex]{0pt}{3.5ex}  Keywords & 10 pt.  \\
% \hline
% \rule[-1ex]{0pt}{3.5ex}  Section heading & 12 pt., bold, left justified  \\
% \hline
% \rule[-1ex]{0pt}{3.5ex}  Subsection heading & 12 pt., italic, left justified  \\
% \hline
% \rule[-1ex]{0pt}{3.5ex}  Sub-subsection heading & 11 pt., italic, left justified  \\
% \hline
% \rule[-1ex]{0pt}{3.5ex}  Normal text & 12 pt. \\
% \hline
% \rule[-1ex]{0pt}{3.5ex}  Figure and table captions &  10 pt. \\
% \hline 
% \end{tabular}
% \end{center}
% \end{table} 

\section{Methods}

\subsection{PRIMAger}
The PRIMAger instrument which contains a hyperspectral imager spanning two bands, PHI1 and PHI2, between 25 and 85 $\mu$m with a spectral resolution of R$\sim$10. Additionally there are four filters in the polarimeter imager between 96 - 235 $\mu$m.  In this work we represent each band of the hyperspectral imager with six photometric filters \cite{Bisigello2024, Donnellan2024} following previous work, where each has a top hat profile with central wavelength and width given in Table \ref{tab:PRIMager}. We show the 5$\sigma$ sensitivities for a survey of 1500h for an area of 1 square degree.  

\begin{table}
\centering
  \caption{PRIMAger filter information and sensitivities. The sensitivity is given as the 5$\sigma$ flux for a survey of 1500h/deg$^2$. The classical confusion limits and achieved sensitivities via Bayesian de-blending techniques are from Donnellan et al. 2024 \cite{Donnellan2024}.}
  \label{tab:PRIMager}
    \def\arraystretch{1.2}%  1 is the default, change whatever you need
    \setlength{\tabcolsep}{4pt}
    \begin{threeparttable}
  \begin{tabular}{cccccc}
  
    \hline

    Filter & Central Wavelength & Filter Width & 5$\sigma$ Sensitivity &  Confusion Limit & \textsc{XID+}, Deep Prior \\
    & $\mu$m & $\mu$m & $\mu$Jy & $\mu$Jy & $\mu$Jy \\
    \hline
PHI1\textunderscore1 & $25.0$ & $2.5$ & $75.1$ & 20 & 82\\
PHI1\textunderscore2 & $27.8$ & $2.8$ & $85.6$ & 27 & 86\\
PHI1\textunderscore3 & $30.9$ & $3.1$ & $96.0$ & 37 & 93 \\
PHI1\textunderscore4 & $34.3$ & $3.4$ & $106.5$ & 51 & 108\\
PHI1\textunderscore5 & $38.1$ & $3.8$ & $116.9$ & 71 & 116\\
PHI1\textunderscore6 & $42.6$ & $4.3$ & $127.4$ & 107 & 149\\
PHI2\textunderscore1 & $47.4$ & $4.7$ & $140.4$ & 161 & 95\\
PHI2\textunderscore2 & $52.3$ & $5.2$ & $160.0$ & 249 & 117\\
PHI2\textunderscore3 & $58.1$ & $5.8$ & $179.6$ & 401 & 138\\
PHI2\textunderscore4 & $64.5$ & $6.5$ & $199.2$ & 667 & 167 \\
PHI2\textunderscore5 & $71.7$ & $7.2$ & $218.8$ & 1120 & 229\\
PHI2\textunderscore6 & $79.7$ & $8.0$ & $238.4$ & 1850 & 285\\
PPI1 & $96$ & $23$ & $89.8$ & 4250 & 281\\
PPI2 & $126$ & $15$ & $117.6$ & 12300 & 747\\
PPI3 & $172$ & $20.3$ & $160.8$ & 28400 & 2650\\
PPI4 & $235$ & $27.6$ & $187.0$ & 46000 & 7030\\
    \hline
  
  \end{tabular}

  \end{threeparttable}
 \end{table}

Considering PRIMA's 1.8m mirror and the wavelength coverage, confusion will naturally be a concern for source identification. We refer the reader to Donnellan et al. 2024 \cite{Donnellan2024} for details on how best to mitigate this problem. In Table \ref{tab:PRIMager}, we report the sensitivities corresponding to the classical confusion limit as well as those based on mitigating techniques. In particular, the use of \textsc{XID+} from Hurley et al. 2017 \cite{Hurley2017}, which uses a Bayesian framework to de-blend sources using some prior. We report the sensitivities using the deep catalog prior in Table. \ref{tab:PRIMager}. Although the sensitivity is reduced significantly for the longest wavelength filters ($\gtrsim80\mu$m) when considering confusion, the dust continuum is typically bright at these wavelengths (restframe $\gtrsim 12 \mu$m) and so we find that accounting for confusion makes little difference to the total number of objects that can be detected.

\begin{figure}
%\hspace*{-0.5cm}                                           
	\includegraphics[width=\columnwidth]{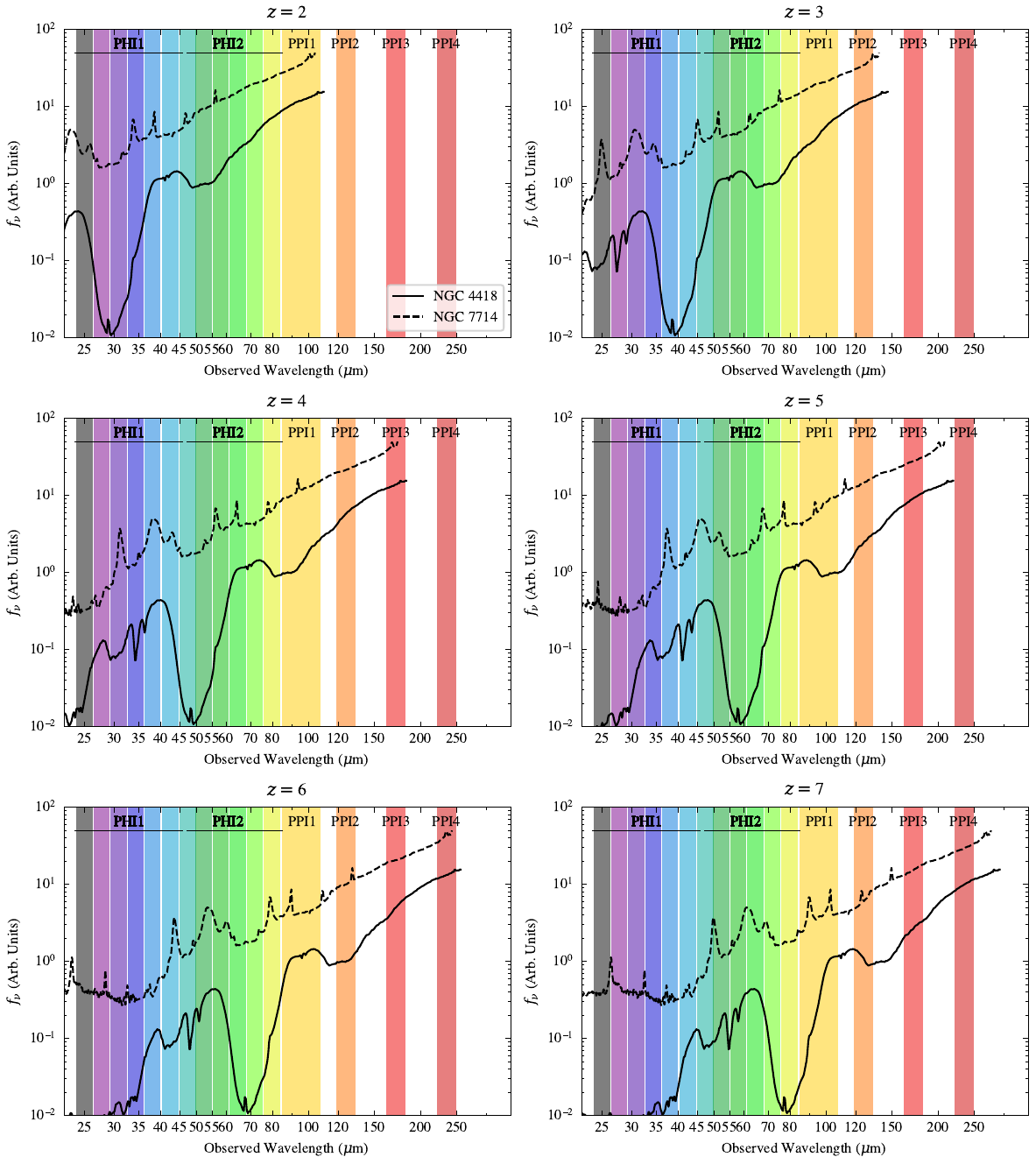}
    \caption{Demonstration of PRIMA's filters at detecting various spectral features, such as the 9.8 $\mu$m silicate absoprtion, at different redshifts. The solid line shows the Spitzer IRS + Akari spectrum of NGC 4418 while the dashed line shows NGC 7714 - a typical star-forming galaxy. 
    The filters, shown as the shaded bands, are described in Table \ref{tab:PRIMager}.}
    \label{fig:SpecWithRedshift} 
\end{figure}

At an R$\sim10$, the resolution at 25 - 85 $\mu$m is sufficient to detect the continuum shape of extremely obscured objects as shown in Fig. \ref{fig:SpecWithRedshift}. In particular the 9.8$\mu$m silicate absorption will be measurable from  $z\sim2-7$, allowing for the selection of obscured nuclei using different combinations of colors \cite{Garcia-Bernete2022} from the filters shown in Table \ref{tab:PRIMager}. We demonstrate this in Fig. \ref{fig:SpecWithRedshift}, where the Spitzer IRS \cite{Spoon2007} + Akari \cite{Imanishi2010} spectrum of NGC 4418 is shown at different redshifts to represent the typical spectrum of a deeply obscured nucleus. We also show the spectrum of NGC 7714 as a typical star-forming galaxy where the spectrum is dominated by PAHs. We use the spectrum from the GALSEDATLAS \cite{Brown2014} which combines the Spitzer and Akari spectrum.

An important consideration for subsequent analysis is well constrained photometric redshifts. 
The presence of the relatively broad PAH features and/or silicate absorption features in combination with the good filter coverage of PRIMAger, is sufficient to simultaneously determine a photometric redshift. A detailed analysis of the effectiveness of the photo-$z$ measurements with PRIMA is beyond the scope of this work and deserves a dedicated analysis. However, previous studies with AKARI, \cite{Negrello2009} have found that mid-IR determined photometric redshifts are recovered to an accuracy within $\sim10\%$. Considering the wider filters of AKARI compared to PRIMA, suggests photometric redshifts can accurately constrained to within $\sim10\%$. Moreover, the uncertainty of this measurement is likely smaller than the width of the redshift bins we use in our color selection (see section \ref{sec:PRIMAgerResults}).

\subsubsection{Galaxy Samples}
In this work we demonstrate PRIMA's ability to identify and characterize deeply obscured nuclei at $2 < z < 7$. To do this we use archival Spitzer spectra of local (U)LIRGs in the IDEOS database \footnote{\url{http://ideos.astro.cornell.edu/}} \cite{IDEOS} and follow the same methodology as Donnan et al. 2023a \cite{Donnan2022}.

We use the \textit{HERschel} Ultra luminous infrared galaxy Survey (HERUS) sample \cite{Farrah2013} to obtain a sample of 42 local ULIRGs. Additionally, we use the Great Observatories All-sky LIRG Survey (GOALS) \cite{Armus2009} survey which contains 179 LIRGs and 22 ULIRGs. Out of the full sample, we use those that have Spitzer IRS data and thus we have 143 LIRGs in our sample. Finally, we use a pure star-forming sample \cite{Hernan-Caballero2020}, which consists of 106 galaxies.

\subsubsection{\textsc{SPRITZ} Simulation}
\label{sec:Simulation}
We use the the Spectro-Photometric Realisations of IR-Selected Targets at all-$z$ (\textsc{SPRITZ}) \footnote{\url{http://spritz.oas.inaf.it/}} simulation \cite{Bisigello2021} which provides predictions of the number of galaxies at different redshifts, following the work of Bisigello et al. 2024 \cite{Bisigello2024}. This simulation is empirically driven, with the population of galaxies being generated based on observed luminosity functions of different galaxy types, namely, star-forming (SF), AGN, composite and passive. The proportion of each galaxy type for a given luminosity and redshift is selected from observed IR luminosity functions \cite{Gruppioni2013}. Each type has an associated SED template assigned depending on the properties of each object, however we are cautious to not rely on the templates as they do not properly account for deeply obscured nuclei. For more details see Bisigello et al. 2021,2024 \cite{Bisigello2024, Bisigello2021}.

\subsection{FIRESS}
% Desxribe population of objects in LIR/redshift bins from the imaging
% Take a representative sample with the spectrograph at R~100
The FIRESS instrument is a spectrograph, observing between 23 - 235 $\mu$m with a spectral resolution of R$\sim$100. With FIRESS, a 5$\sigma$ line flux of $3\times10^{-19}$ Wm$^{-2}$ can be obtained with 1 hour of integration. This corresponds to a flux level of $\sim$ 100 $\mu$Jy which is sufficient to obtain quality spectra of a significant number of candidates detected in the PRIMager surveys. Increasing the exposure time to 5 hours yields a depth of $\sim$45 $\mu$Jy. To test the ability of FIRESS to measure the spectra of deeply obscured nuclei, we simulate a number of objects of varying luminosity at different redshifts. We use the spectrum of Arp 220, which hosts a deeply obscured nucleus \cite{Falstad2021} as well as circumnuclear star-formation.

\begin{figure}
%\hspace*{-0.5cm}                                           
	\includegraphics[width=\columnwidth]{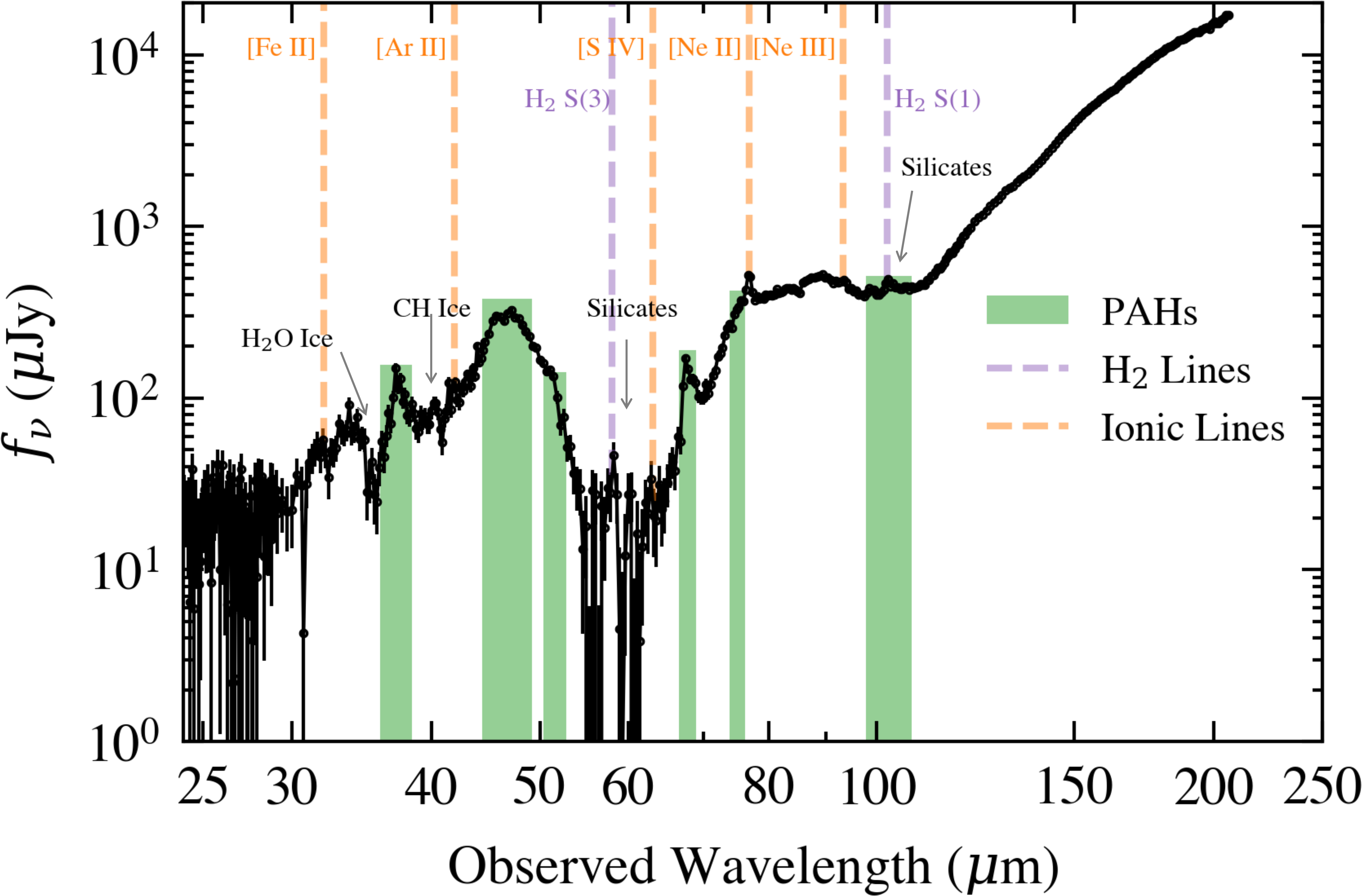}
    \caption{Example spectra as observed by FIRESS of an obscured nucleus within a HLIRG at $z=5$ after 5 hours of integration. We use the spectrum of Arp 220 as a template scaled to an infrared luminosity of $10^{13.15} L_{\odot}$ at $z=5$. Key spectral features are labelled.}
    \label{fig:Spectra} 
\end{figure}

To simulate the spectra, we scale the Spitzer spectra to match the photometry of objects in the \textsc{SPRITZ} simulation that are identified as deeply obscured nuclei (see Section \ref{sec:PRIMAgerResults}). We show the resulting simulated spectra of an obscured nucleus at $z=5$, in Fig. \ref{fig:Spectra}. We find that the continuum and emission/absorption features are present, allowing the recovery of PAH fluxes, ionic line fluxes such as [Ne II], and absorption from H$_2$O ice at restframe $\sim 6 \mu$m. The spectra is therefore of a sufficiently high quality to characterize the nature of deeply obscured nuclei candidates identified from the imaging surveys.

% \subsection{Selecting Highly Obscured Nuclei}

% %Subsection with spectra Similar to Spitzer
% % Techniques such as PAHDecomp to measure PAHs 

% % Can we use torus models

\section{Results}

\subsection{The Population of Deeply Obscured Nuclei at $2<z<7$.}
% %Show how different galaxy types appear with the imager as a function of redshift - show selection for different redshift bins

\subsubsection{PRIMAger}
\label{sec:PRIMAgerResults}
To identify deeply obscured nuclei in the PRIMAger survey, we select filters from Table \ref{tab:PRIMager}, that are sensitive to the silicate absorption at 9.8$\mu$m at a given redshift. While there is enough wavelength range to go beyond $z=7$, the sensitivity limits detecting a reasonable number of targets as discussed in section \ref{sec:Simulation}. For each redshift slice we select a combination of three filters. One of them covers the silicate absorption band \cite{Garcia-Bernete2022} while the other two are selected to measure continuum outside the silicate band. In particular we choose one to be at $\sim13 \mu$m restframe just outside the silicate absorption and the other to be at $\sim30 \mu$m restframe, which is sensitive to the relatively colder dust. This arrangement helps to discriminate against type 2 AGN, which may have a deep silicate feature but show a hotter far-IR continuum \cite{Garcia-Bernete2019}. It is worth noting that other filter combinations can also be used, such as using the 3-5$\mu$m slope as was presented by Garcia-Bernete et al. submitted.

In addition, due to the good filter coverage, PRIMA will be able to sample the SED well (see section \ref{sec:Torus}), and so one can both determine a photometric redshift and classify the nature of the object purely via SED fitting without any prior color selection. In this work we choose to show the color section as it clearly illustrates the strength of the silicate absorption for identify deeply obscured nuclei and can be quantified, without relying on specific SED fitting codes. 

%We avoid presenting such colors in this work as the Spitzer spectra in our samples do not contain any data at wavelengths below 5 $\mu$m, however this would be ideal for the higher-$z$ targets, as the long wavelength filters may suffer from confusion \cite{Donnellan2024}. 

\begin{figure}
\hspace*{1.5cm}                                           
	\includegraphics[width=13cm]{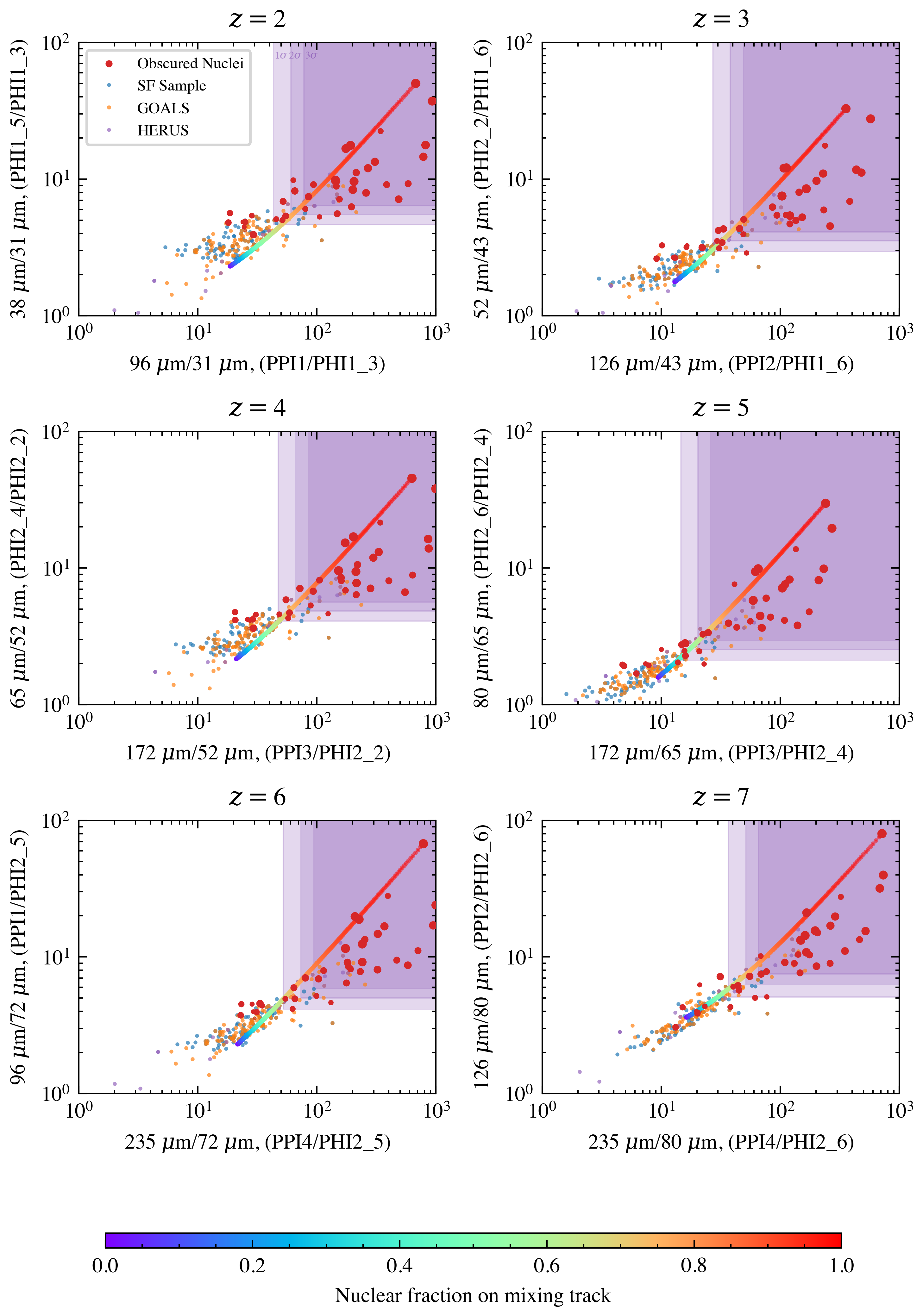}
    \caption{Color selection for identifying deeply obscured nuclei with PRIMAger. Each panel shows a different redshift slice with varying filter combinations. The blue dots show the pure star-forming sample \cite{Hernan-Caballero2020}, the orange show the GOALS \cite{Armus2009} sample while the purple points show the HERUS sample \cite{Farrah2013}. The red points show objects identified as hosting a deeply obscured nucleus in the GOALS and HERUS samples from Spitzer spectra \cite{Donnan2022}, where the size of the points reflect the nuclear fraction as measured in Donnan et al. 2023a \cite{Donnan2022}. The purple box displays the selection criteria for deeply obscured nuclei. This is defined based on the mean and standard deviation, $\sigma$ of the pure star-forming sample, where three boxes are shows as 1$\sigma$, 2$\sigma$, 3$\sigma$ from this mean. The colorbar shows a track for a simulated spectrum consisting of a nuclear component, NGC 4418, and a star-forming component, NGC 7714. The color indicates an increasing nuclear fraction to the total SED. }
    \label{fig:ColorSelection} 
\end{figure}

Fig. \ref{fig:ColorSelection} shows the color measurements for the star-forming \cite{Hernan-Caballero2020}, HERUS and GOALS samples for different redshift slices. The red dots represent objects identified as hosting deeply obscured nuclei as per Donnan et al. 2023a \cite{Donnan2022}.

Donnan et al. 2023a \cite{Donnan2022} provided two criteria to select deeply obscured nuclei. The first was based on PAH EW ratios \cite{Garcia-Bernete2022} while the second was based on modeling the spectra to infer the optical depth of the nuclear continuum. In this work we use the second classification. For a comparison between the two see Donnan et al. 2023a \cite{Donnan2022}.

% which predicted $\sim$ 30 $\%$ of ULIRGs and $\sim$7 $\%$ of LIRGs hosted deeply obscured nuclei. The second was based on modeling the spectra to infer the optical depth of the nuclear continuum which predicted $\sim$ 40 $\%$ of ULIRGs and $\sim$20 $\%$ of LIRGs hosted deeply obscured nuclei. The latter method was more consistent with the CONquest results using HCN-vib emission in the sub-mm \cite{Falstad2021}. This discrepancy was attributed to dilution by the host galaxy, and so the modeling method provided more objects that suffered from dilution by circumnuclear star-formation. In this work we use this selection to generate the obscured sample as it will provide more conservative estimates of PRIMA's performance.
Using the Donnan et al. 2023a \cite{Donnan2022} selected deeply obscured nuclei as the ground truth, we create a selection criteria for PRIMA using the pure star-forming sample \cite{Hernan-Caballero2020}. To generate a color selection for deeply obscured nuclei, we first measure the mean color and standard deviation of the star-forming sample. We then create the deeply obscured nuclei color selection as being  1$\sigma$, 2$\sigma$ and 3$\sigma$ greater than the star-forming mean. This is shown in Fig. \ref{fig:ColorSelection}.

We find that the color selection does a good job at identifying the most deeply obscured objects however the method may miss objects where the emission from the host galaxy is very strong. This is indicated by the size of the points plotted in Fig. \ref{fig:ColorSelection}, where the smaller the marker, the lower the nuclear fraction as measured in Donnan et al. 2023 \cite{Donnan2022}. The nuclear fraction is defined as the fractional contribution to the mid-IR flux (between 5 $\mu$m and 14 $\mu$m) from the nuclear component using \textsc{PAHDecomp} \footnote{\url{https://github.com/FergusDonnan/PAHDecomp}}. Those objects with the highest nuclear fraction ($\gtrsim60\%$) appear firmly within the selection criteria. We find that the 1$\sigma$ criteria selects 80$\%$ of the total obscured nuclei while the 2$\sigma$ and 3$\sigma$ criteria find 65 $\%$ and 60 $\%$ respectively. While the 1$\sigma$ criteria finds a majority of the obscured nuclei up to $\sim$ 45 $\%$ of the targets selected are ULIRGs and LIRGs that do not host deeply obscured nuclei. This drops to zero with a strict color cutoff but would limit to only the most extreme objects being selected. The 2$\sigma$ limit provides a reasonable selection as the number of false positives counteracts the number of objects missed and so if one is simply interested in the total fraction of deeply obscured nuclei for a given luminosity and/or redshift bin, the fraction of obscured nuclei is reasonably accurate.

The effect of contamination by the host galaxy due to the size of the PRIMA primary mirror and therefore the afforded spatial resolution, will limit the completeness of obscured nuclei that PRIMA can detect. This issue will impact the detection of deeply obscured nuclei in the high$-z$ universe however, PRIMA will be able to detect a large number of deeply obscured candidates, more than any other current or future space/ground facility.

\begin{figure}
%\hspace*{-0.5cm}                                           
	\includegraphics[width=\columnwidth]{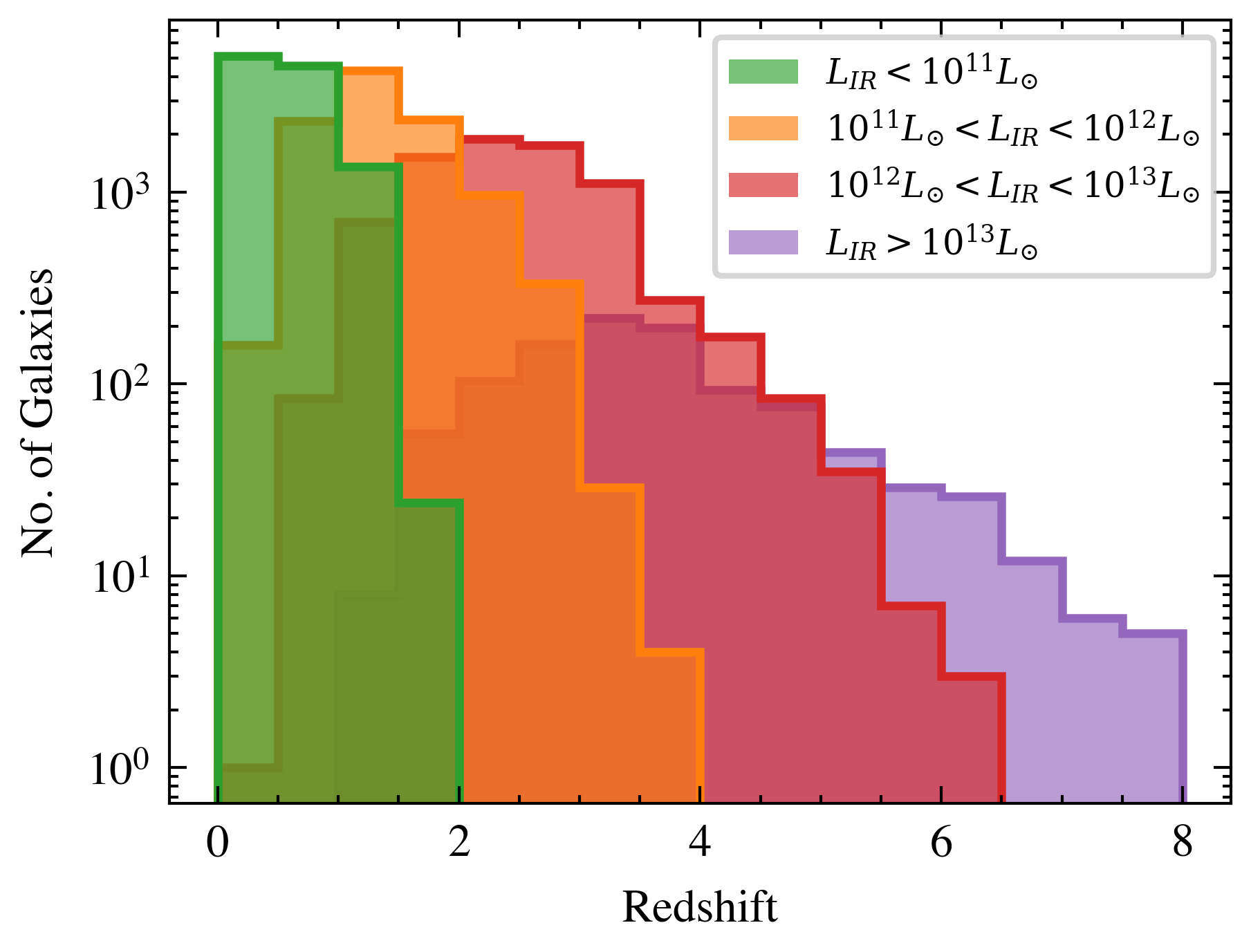}
    \caption{Number of galaxies of varying infrared luminosity that can be detected by PRIMAger using the output of the \textsc{SPRITZ} simulation. These values are for a 1500h/deg$^2$, accounting for confusion using the values shown in Table \ref{tab:PRIMager}.}
    \label{fig:Hist} 
\end{figure}

To test how many galaxies PRIMA can detect, we use the \textsc{SPRITZ} simulation results for a survey of 1500 hours over 1 square degree. We follow Bisigello et al. 2024 \cite{Bisigello2024} and consider a galaxy detected if at least 6 out of the 12 photometric bands have a flux greater than 5$\sigma$, where $\sigma$ is the noise at a given wavelength. We plot the number of detected galaxies in Fig. \ref{fig:Hist}, where we find a reasonable number of galaxies detected out to $z\sim7$. In particular, we find $\sim$9000 galaxies at $z = 1.5 - 2.5$, which lowers to 53 at $z=6.5 - 7.5$. At the highest redshift bin only hyper-luminous infrared galaxies (HLRIGs, $L_{\rm IR} > 10^{13} L_{\odot}$) are bright enough to be detected. 

\begin{figure}
%\hspace*{-0.5cm}                                           
	\includegraphics[width=\columnwidth]{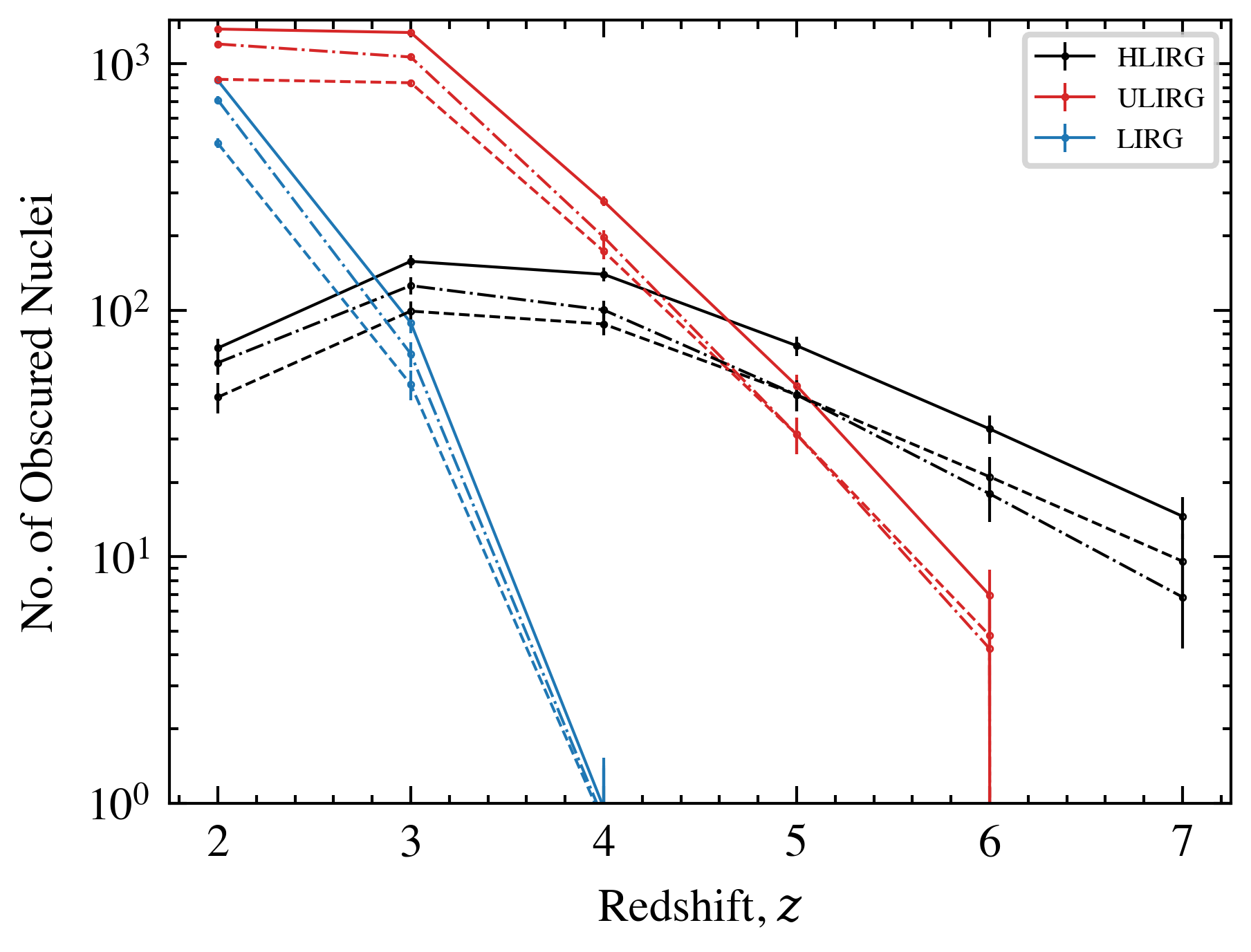}
    \caption{The number of deeply obscured nuclei that PRIMA can detect as a function of wavelength and infrared luminosity. HLIRGs are shown in black, ULIRGs in red and LIRGs in blue. The solid line shows an assumed fraction of sources consistent with local galaxies. The dashed line shows half the fraction of local galaxies. The dashed dot line shows and evolving fraction of obscured sources that decreases from the local value to zero at $z=10$. This has the form, $0.4\left( 1 - \frac{z}{10}\right)$, starting at 0.4 at $z=0$.}
    \label{fig:NoOfSources} 
\end{figure}

Predicting the exact number of galaxies that host deeply obscured nuclei is challenging, as the fraction of HLIRGs, ULIRGs and LIRGs that host such nuclei is not well constrained beyond the local universe. Indeed, PRIMA is well suited to determine the fraction of deeply obscured nuclei in the universe and therefore measure this fraction. We therefore provide a range of predictions depending on the assumed intrinsic fraction, to provide the reader with a general idea of how many objects PRIMA will be able to detect. We plot these predictions in Fig. \ref{fig:NoOfSources} for three different assumptions. For each, we calculate an error on the number of obscured sources, based on a beta distribution \cite{Cameron2011}, which scales with the size of the total sample.

Our first prediction is based on the assumption that $40\%$ of ULIRGs and $20\%$ of LIRGs host deeply obscured nuclei, inline with the local population. As there are no local HLIRGs, we also assume $40\%$ for this population to be conservative but it is likely higher considering the increasing fraction with infrared luminosity. This is shown as the solid lines in Fig. \ref{fig:NoOfSources}. We then assumed half of the local fraction i.e $20\%$ of ULIRGs and HLIRGs, $10\%$ of LIRGs. This is shown as the dashed lines in Fig. \ref{fig:NoOfSources}. We also consider an evolving obscured fraction, which is likely closer to reality. For simplicity we assume a decreasing fraction from local values of $40\%$ of HLIRGs, $40\%$ of ULIRGs, $20\%$ of LIRGs to zero at $z=10$. This is shown as the dashed-dot line in Fig. \ref{fig:NoOfSources}.
% Finally, we show the \textsc{SPRITZ} simulation objects but caution the reader that the templates assumed do not account for deeply obscured nuclei. 

We find that PRIMA will be able to detect $\sim 100-1000$ LIRGs and ULIRGs and $\sim100$ HLIRGs hosting deeply obscured nuclei at cosmic noon. The number of LIRGs drops rapidly, due to their relative faintness, reaching down to zero beyond $z\sim4$. The number of ULIRGs hosting deeply obscured nuclei drops to $\sim10$ at $z\sim6$. Beyond that redshift they are no longer detected. The number of HLIRGs remains $\lesssim10$ out to $z\sim7$. These numbers vary depending on the assumed intrinsic fraction of obscured sources, however even when this fraction is as low at $\sim10\%$ of HLIRGs at $z\sim7$, a non-zero number of sources will be detected. This exercise demonstrates that PRIMA will likely make considerable strides in constraining the number of deeply obscured nuclei up to $z\sim7$.

% We note however that the SEDs chosen in the simulation may lack the appropriate template to match the most obscured nuclei. Indeed, one can see ``tracks'' in the colors plotted in the figure, and in fact many of the objects show the same color and so are plotted on top of each other, which is due to the limited number of templates used to generate the SEDs in the simulation. 

% We measure the colors of the galaxies in Fig. \ref{fig:Hist} in redshift bins, and plot it in Fig. \ref{fig:ColorSelection2}. We find that the number of objects detected are reasonable out to $z\sim7$, with the highest redshift bin containing 13 ULIRGs and 52 HLIRGs, although very few are selected as highly obscured nuclei. This is likely due to the SEDs chosen in the simulation lacking the appropriate template to match the most obscured nuclei. Indeed, one can see ``tracks'' in the colors plotted in the figure, which is due to the limited number of templates used to generate the SEDs in the simulation.

\begin{figure}
%\hspace*{-0.5cm}                                           
	\includegraphics[width=\columnwidth]{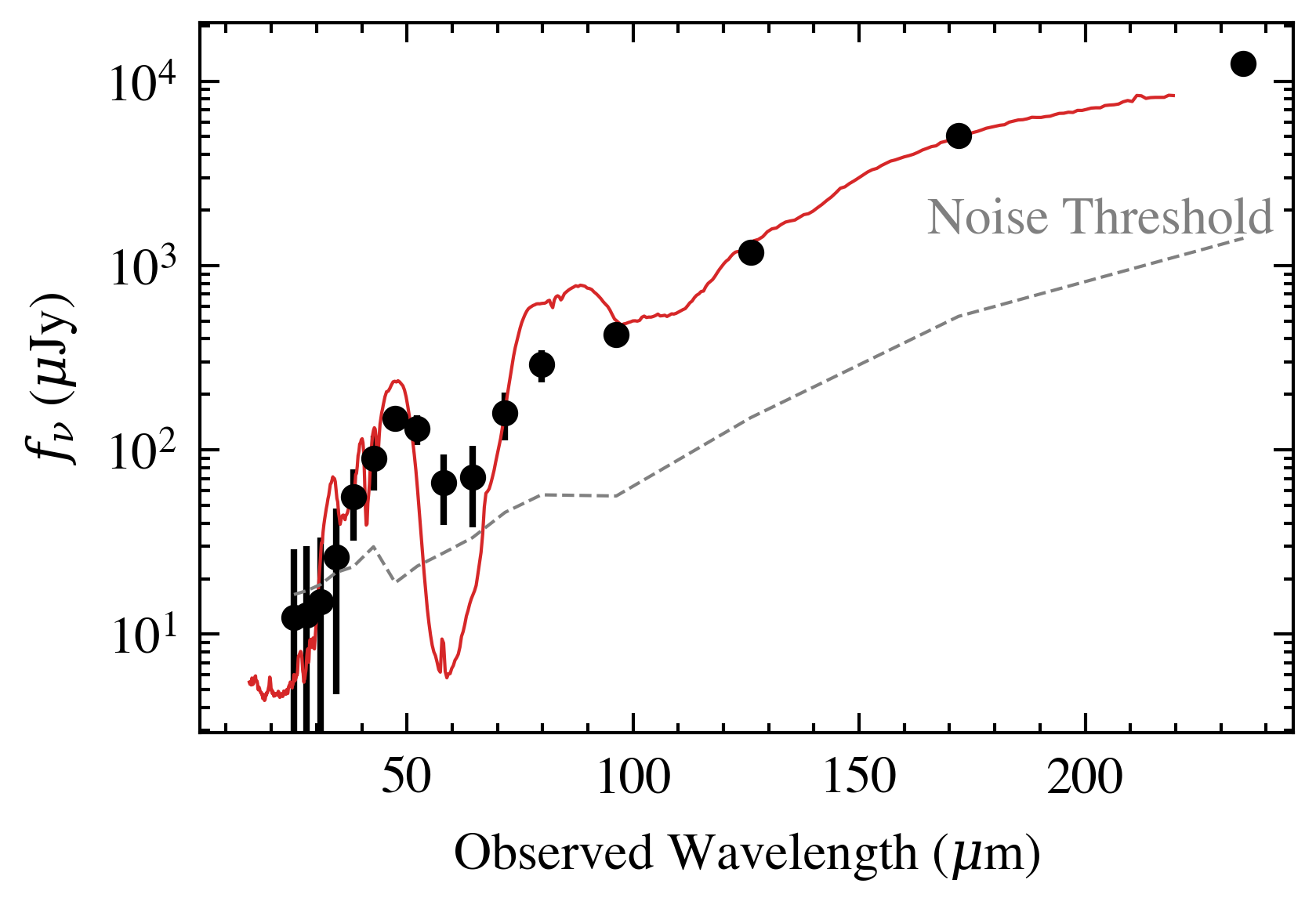}
    \caption{Example of a deeply obscured nucleus at $z\sim5$, selected from the \textsc{SPRITZ} simulation. This particular object has an infrared luminosity of $10^{13.38} L_{\odot}$. The figure shows the 9.8$\mu$m silicate absorption at an observed wavelength of $\sim$60 $\mu$m, where the absorption is sufficiently deep to hit the noise threshold, as shown by the gray dashed line. In comparison is the Spitzer + Akari spectrum of NGC 4418 which has an extremely deep silicate feature - note this is not simulated but purely shown for reference.}
    \label{fig:SED} 
\end{figure}

It is also worth noting that if the silicate absorption feature is sufficiently deep, the flux in the bands covering this feature will drop below the noise threshold of the instrument and thus the measured color becomes a lower limit. We show an example SED of an object showing a very deep silicate feature at $z\sim5$ in Fig. \ref{fig:SED} in comparison to the Spitzer spectrum of NGC 4418, which has one of the deepest silicate features in the local universe. While such objects will still be successfully classified as hosting deeply obscured nuclei, they will not appear in the top right corner of the color-color plot but instead appear as an upper limit within the box.
% The strength of the imaging lies in the total number of galaxies that can be observed however it can suffer from contamination as we have demonstrated. With spectroscopy from FIRESS, we can select objects with a much cleaner selection using the PAH EW method \cite{Garcia-Bernete2022} or by modeling the spectra to measure the nuclear optical depth \cite{Donnan2022}.

\subsubsection{FIRESS}
\label{sec:FIRESS}
% Mention how this can be used to calibrate the imaging selection to get better number statistics 
Photometrically selected candidates can be followed up to confirm their nature using FIRESS spectroscopy. As we demonstrated in Fig. \ref{fig:Spectra}, we can obtain quality spectra within a reasonable integration time. Using spectroscopic observations we will be able to properly understand the nature of these targets. Only through spectroscopy we will be able to confirm the presence of a deeply obscured nucleus and study their properties through detailed modeling of the FIRESS spectra.

As demonstrated in Fig. \ref{fig:Spectra}, the spectra will include a number of important features. Apart from the silicate absorption features, absorption from ices and emission from PAHs are clear. Additionally, many emission lines can be detected, depending on the nature of the target.

Ice features of H$_2$O and CH can be seen at $\sim$36 $\mu$m and 40$\mu$m respectively for a galaxy at $z=5$. These features have been shown to trace the obscuring gas column density well \cite{Garcia-Bernete2024a}, and are commonplace in obscured nuclei and even in some ``typical'' Seyfert AGN. While the role of ices in the dusty structure in galaxy nuclei is not well understood, by the launch of PRIMA, work with JWST data will have advanced our understanding, however this will be confined to $z<3$. With PRIMA, we will be able to trace the feature over cosmic time to understand how ices form and evolve in the nuclei of galaxies.

With a representative sample of targets we can search for signs of evolution on the nuclear structure via high ionization potential lines such as [Ne V] (14.32 $\mu$m restframe). Most obscured nuclei in the local universe show no emission from these lines, even with high sensitivity JWST/MIRI spectra in the local universe \cite{Donnan2022b, Perna2024}, however from the large sample Spitzer selected objects in Donnan et al. 2023a \cite{Donnan2022}, only 1 target showed evidence of [Ne V], IRAS 19254-7245. With the large number of targets that PRIMA can identify, we will be able to see if such high ionization lines are present at high-$z$. Their presence would indicate that these objects are fundamentally different compared to their local counterparts. PRIMA therefore has the ability to measure how these objects evolve over cosmic time.

% This is likely related to some opening in the dusty obscuring structure, allowing these ionizing photons to escape. How these dusty nuclei evolve is a key open question, and by identifying objects that show high ionization potential lines, we may be witnessing this evolution taking place. However the relative rarity of these lines in the local universe suggest we need large number statistics to identify cases where the high ionization potential lines are present. FIRESS is ideal for this task, being able to observe a large number of targets at a variety of redshifts.

%Based on sensitivity/number of objects how well can the number of CONs be measured

%Different properties, with the large number of targets, can we find high-ip lines/when do they switch on?

% \subsection{Synergies with other facilities}
% From the PRIMAger survey we will have numerous (specify from the work in previous sections) galaxies that can be characterized using other facilities. For example, with JWST, we can obtain the host galaxy properties such as the stellar population and morphology while ALMA will probe the cold dust content. As these facilities require targeted observations, this will come in the form of followup observations on specific objects that PRIMA will select, or if there is any overlap with existing surveys. 

\subsection{Torus Modeling}
\label{sec:Torus}
% As mentioned before our current models struggle to fit. 
At present, the majority of  deeply obscured nuclei lack high ionization potential lines and X-ray emission, therefore the infrared SED is the only reliable diagnostic of the properties of the buried AGN. However at high-$z$, there is a lack of understanding of AGN tori leading to torus templates often being blindly applied at high-$z$. As the majority of AGN are obscured at $z>1$ \cite{Hickox2018}, understanding the nature of the dusty obscuring structure is critical to understanding the growth of SMBHs and ultimately how galaxies evolve. With the wavelength coverage PRIMAger and FIRESS, one can fit and refine the torus models to understand the nature of the most obscured nuclei. 

As an example, we fit a candidate selected from the \textsc{SPRITZ} simulation at $z=7$ and fit the photometry with the \textsc{SMART} SED fitting code \cite{Varnava2024}. This tool uses radiative transfer simulations for all the components of the SED including a spheroidal host component and an AGN torus. We use the \textsc{CYGNUS} library of torus models \cite{Efstathiou1995, Efstathiou2013}, to fit this object. 

\begin{figure}
%\hspace*{-0.5cm}                                           
	\includegraphics[width=\columnwidth]{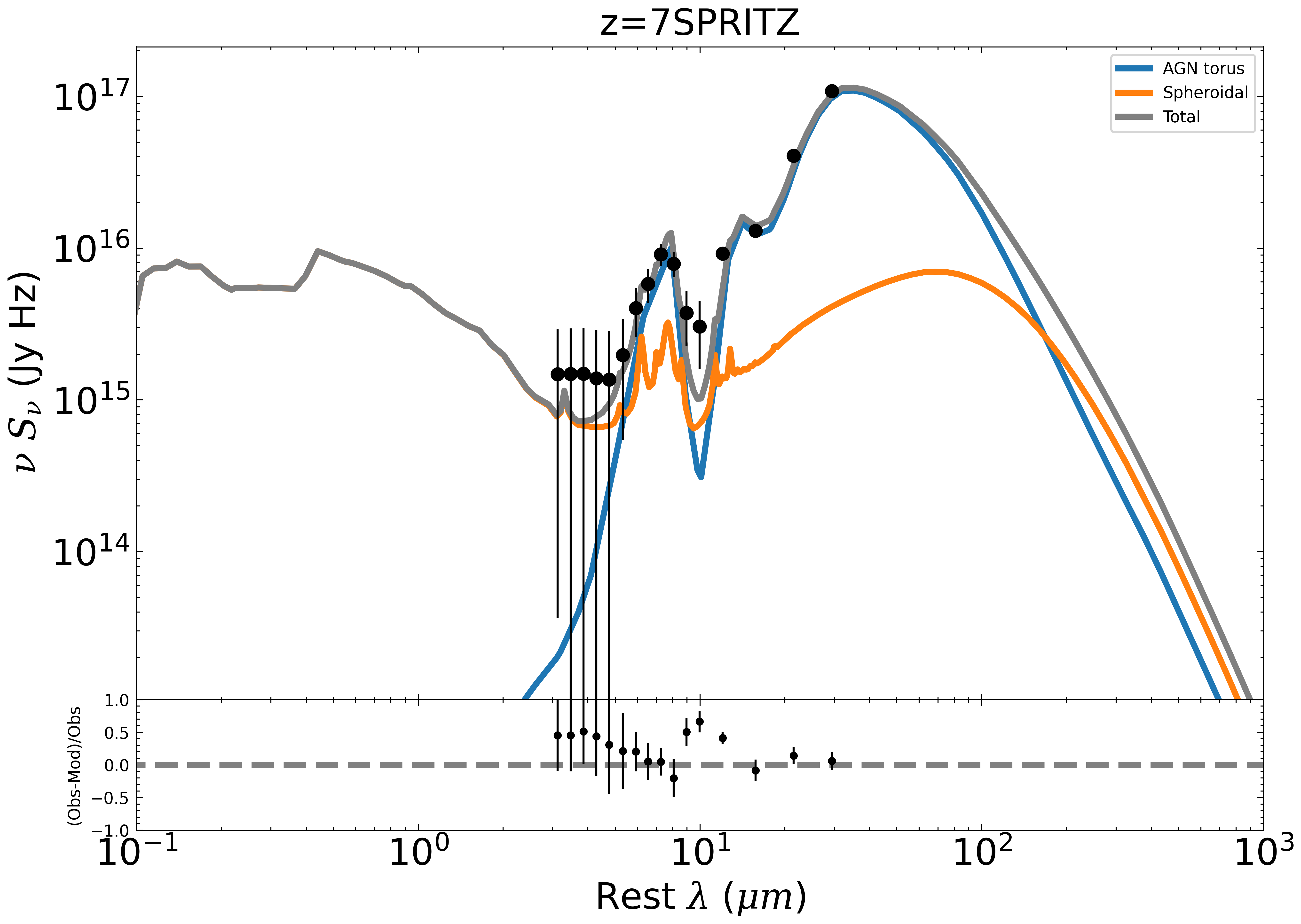}
    \caption{Example of a fit to a HLIRG hosting an obscured nucleus identified from the \textsc{SPRITZ} simulation at $z=7$ using the \textsc{SMART} SED fitting code \cite{Varnava2024}. This particular source has an infrared luminosity of $10^{13.78} L_{\odot}$. }
    \label{fig:SEDFit} 
\end{figure}

The fit is shown in Fig. \ref{fig:SEDFit}. As expected the continuum is dominated by the torus, with the spheroidal host contributing the PAH features. 

Iterating and refining the torus models to fit the obscured nuclei that we will detect is critical for inferring physical parameters of the AGN within. For example, the intrinsic luminosity and thus accretion rate strongly depends on the torus geometry \cite{Efstathiou1995} %anistropy correction
and so for determining the cosmic accretion rate density to map the growth of SMBHs across cosmic time requires such detailed understanding of the torus. 

%\subsection{Evolution of ices}

\section{Conclusions}

We have demonstrated the power of PRIMA to detect and characterize the deeply obscured galaxy nuclei over cosmic time. The main strength of PRIMA is the large sample of galaxies that can be observed in the resframe mid-IR up to $z=7$. We showed that with PRIMAger, the good wavelength coverage and sensitivity allows the successful detection of deeply obscured nuclei between $z=2-7$, where $\sim$65 $\%$ of obscured nuclei can be recovered. Those recovered have the deepest silicate feature, while the remaining $\sim$35 $\%$ are likely to suffer from contamination by the host galaxy. With different assumptions on the intrinsic fraction of sources hosting deeply obscured nuclei, we have shown that PRIMA can effectively constrain this number down to $\sim10\%$ of HLIRGs at $z\sim7$.

With FIRESS, high quality spectra can be obtained, with 5 hours of integration, which enables us to study/constrain the properties of deeply obscured nuclei. Additionally, spectral features such as H$_2$O ice can be studied at high-$z$, opening a new window onto how dust evolves in the early universe. 

We finally showed that the spectral coverage of PRIMAger is sufficient to constrain torus models, even out to $z=7$, which will revolutionize our understanding of the evolution of dust and its role in AGN obscuration.  With PRIMA we will therefore be able to construct a complete census of the most obscured galaxy nuclei over cosmic time and therefore reveal the hidden growth of SMBHs.

\subsection*{Disclosures}
No Conflicts of Interest

\subsection* {Code, Data, and Materials Availability} 
The Spitzer IRS spectra used in this work are available from the IDEOS database \footnote{\url{http://ideos.astro.cornell.edu/}}.

\subsection* {Acknowledgments}
FRD acknowledges support from STFC through grant ST/W507726/1. DR and IGB acknowledge support from STFC through grant ST/S000488/1 and ST/W000903/1. We thank the referees for their careful review.

%%%%% References %%%%%

\bibliography{report}   % bibliography data in report.bib
\bibliographystyle{spiejour}   % makes bibtex use spiejour.bst

%%%%% Biographies of authors %%%%%
\vspace{2ex}\noindent\textbf{Fergus R. Donnan} is a PhD student at the University of Oxford, whose research focuses on star-formation and AGN activity in dusty galaxies, primary in the mid-infrared, to study the growth of the most obscured black holes.

\vspace{2ex}\noindent\textbf{Dimitra Rigopoulou} is a Professor at the University of Oxford, whose research interests are the evolution of galaxies through multi-wavelength, in particular mid-infrared/sub-mm. In particular much of her research focuses on the dust obscured universe.

\vspace{2ex}\noindent\textbf{Ismael Garc{\'i}a-Bernete} is a Atracción de Talento Investigador “César Nombela” Fellow at CAB, Madrid, researching the role of AGN in galaxy evolution, with a focus on dusty tori, PAHs, molecules etc.

\vspace{2ex}\noindent\textbf{Laura Bisigello} is a post-doc researcher at the INAF-OAPd in Padova. Her research focuses on galaxy formation/evolution across cosmic time as well as predictions for future telescopes.

\vspace{2ex}\noindent\textbf{Susanne Aalto} is a professor and Deputy CEO/Deputy President at Chalmers university whose research focus has been radio/sub-mm astronomy studying star-formation, AGN and winds/outflows with a focus on molecular gas. 
\newpage

\vspace{1ex}
%\noindent Biographies and photographs of the other authors are not available.
\listoftables

\listoffigures

\end{spacing}
\end{document}